\documentclass{aa}
\usepackage{txfonts}
\usepackage{graphicx}

\newcommand{\de}{\delta}

\begin{document}

\title{A study of Complexity in Gamma Ray Burst using the\\
	 Diffusion Entropy Approach}
\titlerunning{GRB and the Diffusion Entropy Approach}
\authorrunning{Omodei et al.}

\author{N. Omodei\inst{1,2}
\and J. Bellazzini\inst{3}
\and S. Montangero\inst{4}}
\institute{Dipartimento di Fisica, Universit\'a di Siena,
	 via Roma 56, 53100 Siena \and
	INFN Sezione Pisa, Polo Fibonacci, via F. Buonarroti 2, 56100 Pisa \and
	Dipartimento di Ingegneria Aerospaziale, Universit\'a di Pisa, via Caruso, 56100 Pisa \and
	Scuola Normale Superiore \& NEST-INFM, Piazza dei Cavalieri 7, 56100 Pisa}

\offprints{N. Omodei, \email{nicola.omodei@pi.infn.it}}
\date{Received  / Accepted}
\abstract{
The Diffusion Entropy algorithm is a method that allows the study of correlated non-stationary time series and allows the discrimination between signal and uncorrelated noise.
DE provides a quantitative measure of the {\it complexity}  by means of a scaling index $\de$.
The aim of this paper is to apply this method to study and statistically characterize Gamma-Ray Burst  light curves and to introduce a method to constrain and test GRB models.
}
\maketitle

\section{Introduction}

Gamma-Ray Bursts (GRB), discovered in 1973 by the Vela satellites (Klebesadel et al. \cite{kle73}), are among the most powerful phenomena in the universe. Their typical emission is of the order of $10^{52}$ erg and the nature of these events seems to be destructive, maybe related to explosions. The energy emission during the flaring phase (prompt emission) is in the gamma-ray range, with a $\nu F_{\nu}$ peaked around few hundreds of keV. 

The GRB durations are distributed following a bimodal distribution covering three decades (Kouvelietou et al. \cite{kou93}). Short bursts last less than milliseconds while the longest ones can last up to hundreds of seconds.
Moreover, GRB temporal series (light curves) are typically very different from one burst to another. 
The light curves have many peaks, and in spite of extensive statistical studies, the temporal behavior of GRB remains a puzzle.
Contrary to the very diverse behavior in the time domain, GRB have a simple behavior in the Fourier domain: the Power Density Spectrum, averaged over many bursts, shows a universal power law decay $1/f^\beta$ (with a break at $\approx 1 Hz$) with $\beta\sim 5/3$ (Beloborodov et al. \cite{bel00}).
This characteristic decay has been observed in many other complex systems such as fully developed turbulence in a fluid flow (Frisch \cite{fri96}), self-organized criticality (Bak et al. \cite{bak87}), and, more generally, in systems with long-range correlations. 
One of the methods used to study the ``complexity'' of these systems is Diffusion Entropy (DE), introduced by Scafetta et al. (\cite{scaf01}).
In our analysis we make use of the available light curves with 64 ms time resolution obtained by the Burst and Transient Source Experiment (BATSE) in the four Large Area Detector (LAD) energy channels.

The paper is organized as follows: in Sect. 2 we describe the method, in Sect. 3 we consider different analyses by applying the method to the BATSE light curves. 
We first apply the DE method to two GRB that show quite different light curves. We then apply the DE method to the long GRB population (with $T90\ge 2$ seconds), building up the statistics of the scaling indices in the four energy channels. We find a relationship between the scaling index $\de$ and the power measured in GRB emission (fluence). We show, with the support of simple simulations, how this relation represents a key study for constraining models. 
Finally, in order to investigate the temporal evolution of the complexity in the GRB light curve, we study the Diffusion Entropy as a function of time for the two selected bursts of the first analysis.
Our conclusions are drawn in Sect. 4.

\section{Diffusion Entropy Method}

Diffusion Entropy was introduced in Scafetta et al. (\cite{scaf01}) and it has been applied to many different fields and, in astrophysics, to Solar Flares (Grigolini et al. \cite{grigo02}).
It is based on the central limit theorem and allows the correlation of a given sequence to be measured (Feller \cite{fel71}).
In the standard language of the diffusive processes, given a signal
$\big\{x_{i}\big\}_{i=1,N}$ we can think of each $x_i$ as the length of a
jump of a walker moving in a one dimensional space.  
We create a set of many different diffusion {\it trajectories} by means of a moving window of size $\Delta t$ with $1<\Delta t<N$ which is the same as a boxcar smoothing forward in time from point $t$ to any point $t+\Delta t$. 
In practice, we generate $N-\Delta t+1$ trajectories $y_{j}$ considering the sums:
\begin{equation} \label{y} 
y_{j}(\Delta t)=\sum_{i=j}^{j+\Delta t} x_{i} .
\end{equation} 
Each $y_{j}(\Delta t)$ can be thought of as the final position of the walker after $\Delta t$ time steps. 
The normalized histogram that contains the final positions at time $\Delta t$ ($y_i(\Delta t)$) is the probability distribution $p(y,\Delta t)$ of the final positions $y$ of the walker at time $\Delta t$ for which the Shannon
entropy is:
\begin{equation} \label{shannonentropy} 
S(\Delta t)=-\int_{-\infty}^{+\infty}p(y,\Delta t)\ln(p(y,\Delta t))dy.
\end{equation} 
From the central limit theorem, if the time series $\big\{x_{i}\big\}_{i=1,N}$ has finite variance and is uncorrelated, the probability $p(y,\Delta t)$ satisfies the scaling condition: 
\begin{equation} \label{scaling} 
p(y,\Delta t)=\frac{1}{\Delta t^\de}F(\frac{y}{\Delta t^\de}); 
\end{equation}
then $\de=\de_{\mbox {N}}\equiv0.5$ and $F$ is the Gaussian function for normal (Brownian) diffusion (Feller \cite{fel71}).

More generally, a larger class of diffusive processes can be characterized by the scaling condition (\ref{scaling}) with
$\de\ne\de_{\mbox {N}}$ and non-Gaussian $F$.  
Such processes have been widely observed in nature (Shlesinger et al. \cite{shles93}) and are known collectively as ``anomalous'' processes. For these the central limit theorem hypothesis no longer holds: the signal $x_{i}$ maintains long time correlation (not Markovian) and/or it is not stationary.  
In other words, the DE algorithm smoothes the signal by summing a series of data contained in the window $\Delta t$ and characterizes how the information loss increases with increasing window size.
The rate of information loss for an ``anomalous'' process is faster than for a normal diffusion.
The DE method provides a discrimination between normal and anomalous diffusion by determining the scaling index $\de$.
When the scaling condition (\ref{scaling}) holds, the Shannon entropy $S(\Delta t)$ can be written in the general form 
\begin{equation}
S(\Delta t)=A+\de \ln(\Delta t),
\end{equation} 
independent of the value of $\de$ and of the shape of the function $F$ (Scafetta et al. \cite{scaf01}).

\begin{figure}
\resizebox{\hsize}{!}{
\includegraphics{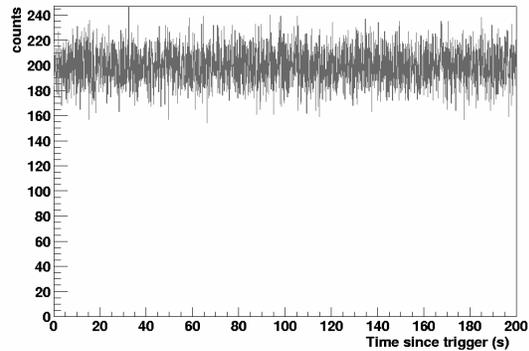}}
\resizebox{\hsize}{!}{
\includegraphics{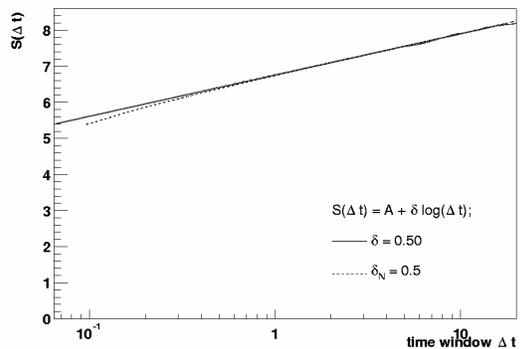}}
\caption{Top: light curve of simulated noise signal. Bottom: diffusion entropy as a function of the temporal window $\Delta t$. (Dashed: $\de=\de_{\mbox {N}}$ theory, Fill: $\de=0.50$ numerical).}
\label{noise_lc_de}
\end{figure}

As a first step we apply DE method to the ``synthetic" light curve, where the counts are distributed according to the Poisson distribution.
We set the binning of the data to 64 ms and we generate 200 seconds of signal. 
As expected from the theory, the $\de$ index is $\de_{\mbox {N}}=0.5$. 
We will refer to this reference signal as ``noise". 
Fig.~\ref{noise_lc_de} shows the noise signal (top) and the corresponding entropy $S(\Delta t)$ as a function of the window $\Delta t$ (bottom).
In the figure each point corresponds to the value of the Shannon entropy $S(\Delta t)$ as a function of the size $\Delta t$ of the window, varying from 64 ms to 20 seconds with steps of 64 ms. 
The choice of this value comes from the bin size of the experimental data and from the duration of the light curve.
We choose the maximum size of the window as $1/10$ of the total duration of the light curve. This condition is required to have enough trajectories $y_i(\Delta t)$ to compute the probability $p(y,\Delta t)$.
However, the peculiar feature of GRB time series, i.e. the short time duration of the prompt emission with respect to the whole data set, makes the analysis more difficult. 
Indeed, the index $\de$ measures the non-stationarity/memory ``intensity'' of the light curve, and, therefore, the diffusion process is governed by the balance between the GRB signal and the background noise. 

Notice that, if we consider the whole BATSE data series, each light curve is composed of several tens of seconds of noise and the GRB signal over-imposed. 
If, for instance, the GRB is short and weak, the measured $\de$ will be close to $\de\sim 0.5$. From this point of view, the DE could mistake a signal with a small contribution of non-stationary behavior for a noisy time series. 
For this reason, we restrict our analysis to the long burst population by selecting only those bursts with $T90\ge 2$ seconds.
This choice ensures that $\de$ indices are uncorrelated with respect to the GRB duration and it also ensures that the measure of the DE is uncorrelated from the experimental acquisition time.


\section{Diffusion Entropy and BATSE data}

We apply the DE method to data available in the CGRO/BATSE archive\footnote{ftp://cossc.gsfc.nasa.gov/compton/data/batse/ascii\_data/64ms/}.
These data are 64 ms binned time series, containing the counts for the four LAD energy channels, 1-4: (1) 20-50 keV, (2) 50-100 keV, (3) 100-300 keV, (4) 300-1000 keV. 
We perform three different studies. 
We first compute the DE of two very different GRB: the scaling indices $\de$ are indeed different from the reference value of noise signals $\de_{\mbox {N}}$.
We compute the DE indices for the entire catalog, and we study the dependence of the $\de$ indices on  the four energy channels. We then study the correlation between the $\de$ indices and the fluence, finding the relationship between the {\it complexity} and the {\it power} in GRB.
We also propose a schematic model to understand this relation.
Finally we show that DE varies during the GRB time evolution.

\subsection{The case of a single burst}

\begin{figure}[ht]
\resizebox{\hsize}{!}{\includegraphics{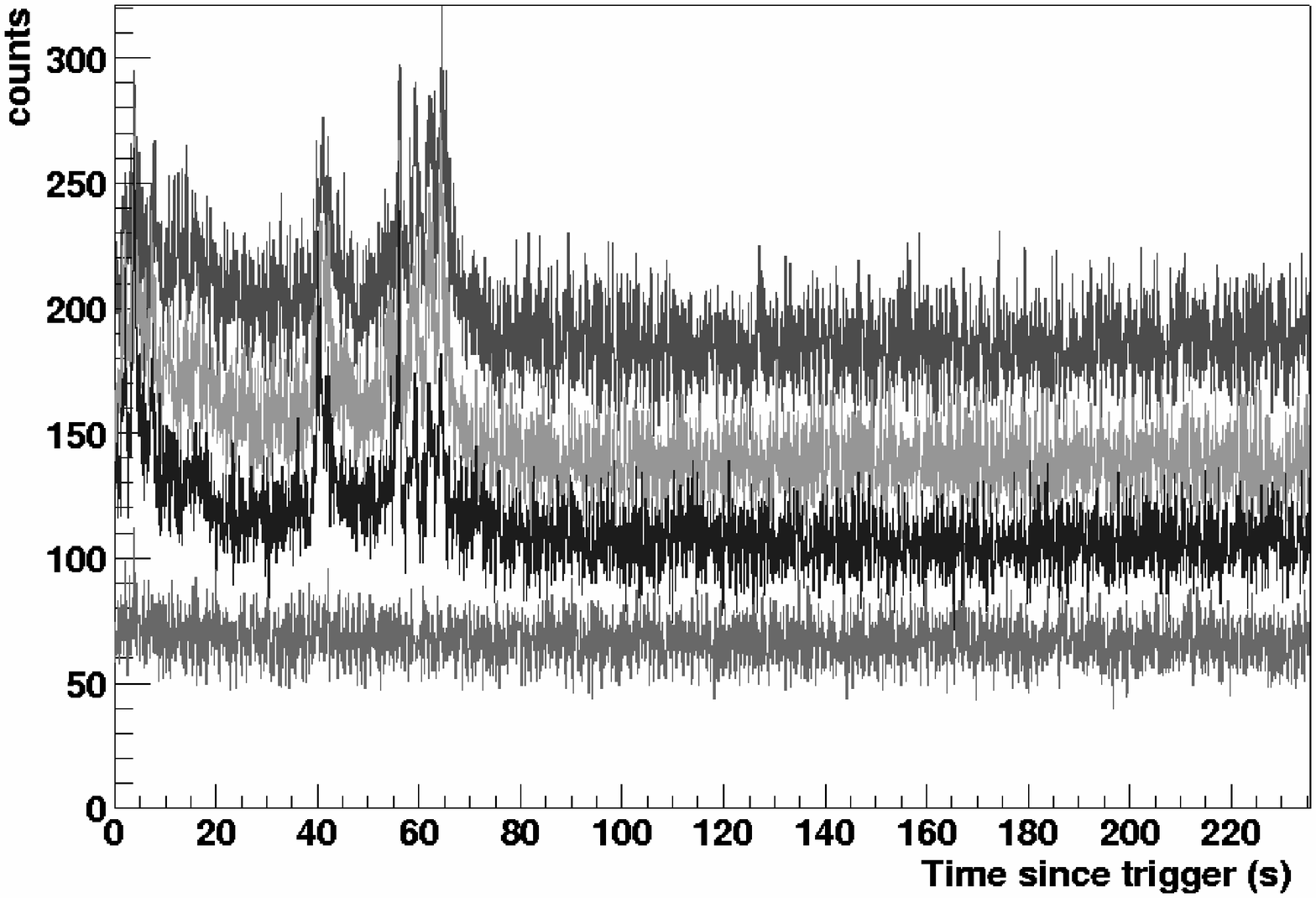}}
\resizebox{\hsize}{!}{\includegraphics{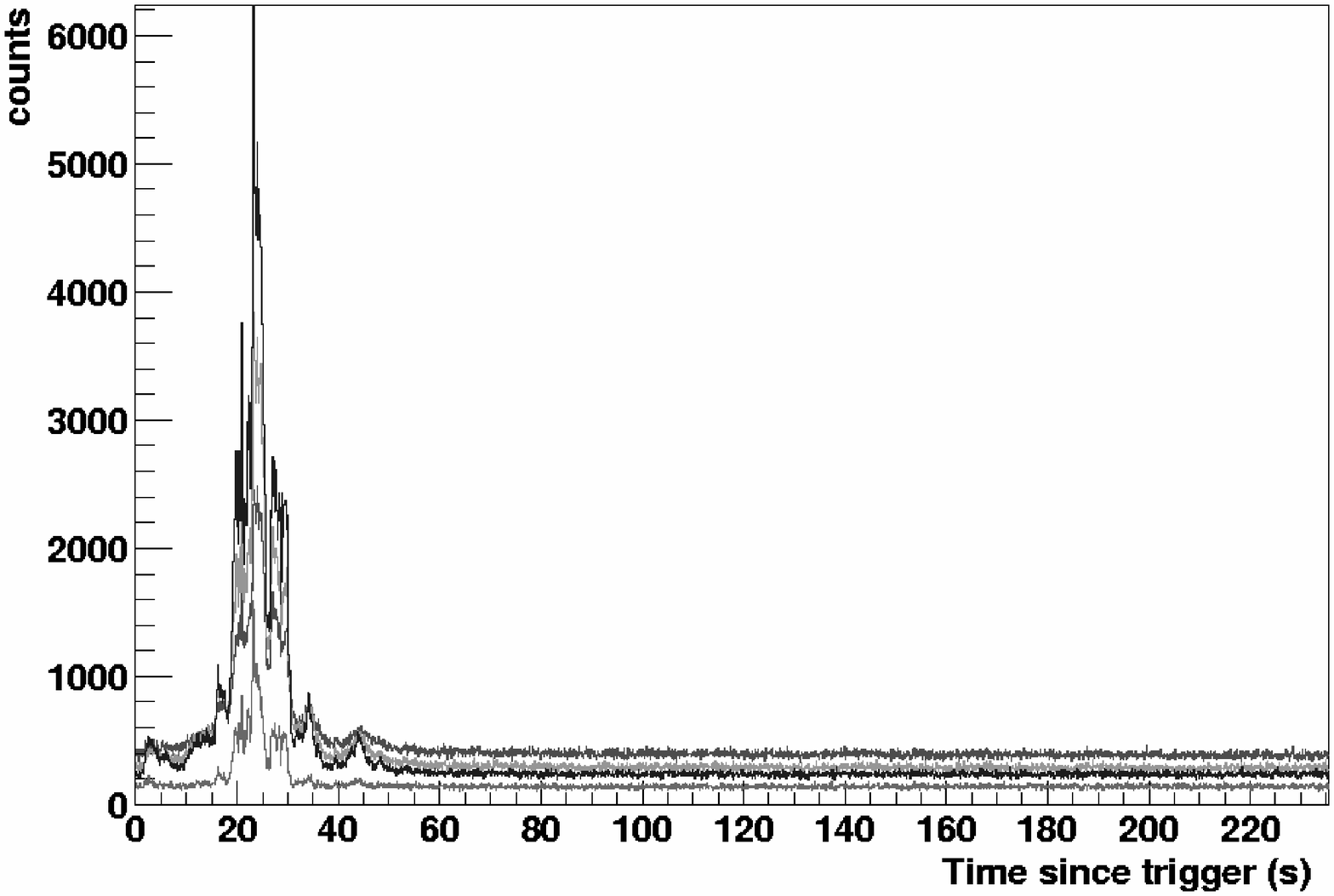}}
\caption{Light curve of the GRB 910429 (top) and GRB 910601 (bottom). Data represents the counts as function of time in four different energy channels. From top to bottom: 20-50 keV, 50-100 keV, 100-300 keV, and 300-1000 keV.}
\label{grb_lc}
\end{figure}

In Fig. ~\ref{grb_lc} we plot the count rate of two different GRB (910429, top and 910601, bottom) as a function of time for each of the four BATSE energy channels. 

 The two selected burst exhibit different characteristics in terms of temporal variability, and peak intensities. The first is a faint, smooth burst while the second is a strong, spiky burst.
We adopt the model that the structure of a GRB can be viewed as the non-stationary increase of the probability of a walker to make longer jumps. 
Since the activity of the GRB lasts only a finite amount of time, this is what we have called {\it non-stationarity}.  
This behavior is evident in the first three channels of GRB 910429 (lowest energies) where the signal exceeds the noise background. 
In the fourth channel the signal seems to be dominated by the noise, and the not stationarity of the signal is less evident.
What we mean by {\it complexity} is essentially the same as anomalous diffusion: it is an
estimate of how far an anomalous diffusive process is from completely uncorrelated noise.
In the case of GRBs the high complexity arises from the high time variability, the nonperiodicity, and the typical non-stationarity of the signal.
Yet our claim is the complexity of a GRB signal derives from the highly structured environment that surround the burst. Indeed, there are reasons to think that the high power released in the GRB
phenomena, combined with the high variability, requires a site where very efficient energy conversion processes take place in a structured, perhaps clumpy, environment with substantial variation from one source to another.
The measurement of the Diffusion Entropy by means of the scaling index $\de$, is indeed an estimation of the complexity of the GRB light curve: the difference between $\de$ and $\de_{\mbox {N}}$ quantifies the ``distance'' between the light curve and the noise.  

\begin{figure}
\resizebox{\hsize}{!}{\includegraphics{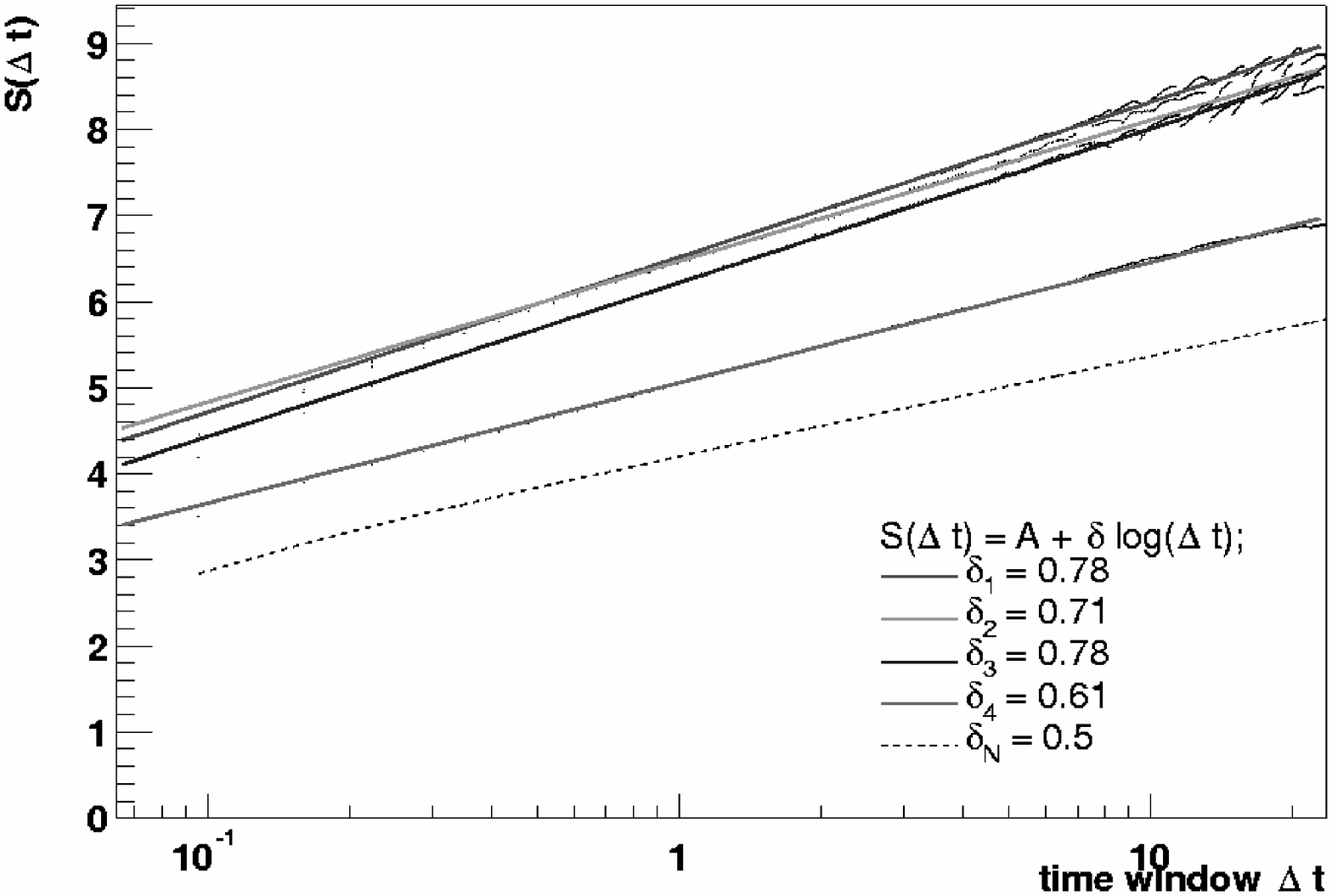}}
\resizebox{\hsize}{!}{\includegraphics{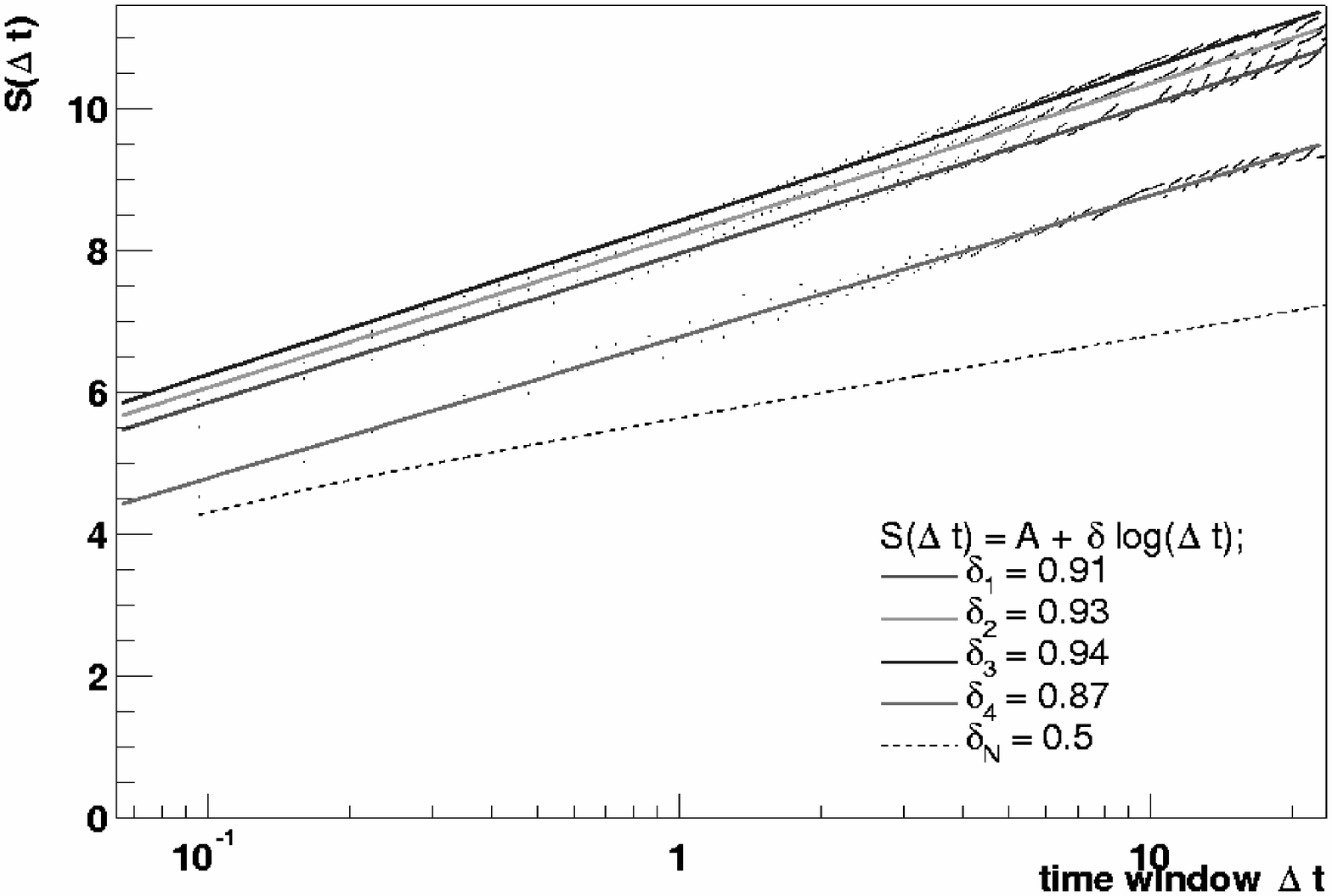}}
\caption{Diffusion Entropy of GRB 910429 (top) and GRB 910601 (bottom), four different channels. Numerical data (dots). Lines: data fits using functions $A+\de_i \log(t)$. The resulting indices $\de_i$ for the channel 1,2,3,4 are in the legends. Dashed line: noise reference value $\de_{\mbox {N}}=0.5$.}
\label{grb_de}
\end{figure}

The results of the application of the DE algorithm to the time series of Fig.~\ref{grb_lc} are shown in Fig.~\ref{grb_de}. 
The lines are the fits to the data with the function $A+\de\ln(\Delta t)$ for the different energy channels. 
The variable $\Delta t$ is the size of the temporal window for which the Shannon entropy $S(\Delta t)$ is evaluated. 
The temporal scale over which the statistics is ``good" is wide since we analyze an order of $N\sim 10^{3}$, corresponding to $\Delta t$ from 64 ms to 24 seconds in 375 steps. 
The resulting indices are shown in Fig.~\ref{grb_de}. GRB 910601 shows higher values in each channel than GRB 910429, reflecting the greater complexity of the former. 

In order to estimate the error on the $\de$ indices due to Poisson fluctuations of photon counting of the original data, we proceeded as follows.
A set of fictitious light curves $\big\{x^{j}_{i}\big\}_{i=1,N; j=1,L}$ was generated from an original $\big\{x^{0}_{i}\big\}_{i=1,N}$ light curve, where $x^{j}_{i}$ is a random number extracted from a Poisson distribution of mean $x^{0}_{i}$.
We find a Gaussian distribution of $\de$ indices. The dispersion of which gives an estimate of the uncertainty in $\de$. 
This analysis with $L=500$ gives an error $\de_{\mbox{err}}\approx 0.05$.
Therefore, studying the $\de$ indices we can discriminate a signal from the noise also in ``critical'' situations, as in GRB 910429 (top) channel 4, where the signal-to-noise ratio is close to one. 
Indeed, in this particular case, $\de=0.61\pm0.03$.
Even if the fourth channel series is dominated by stationary noise, it has a non-stationary component (i.e. a memory effect): the computation of the scaling index $\de$ reveals this ``anomaly" and gives a unique measure of its value.
We notice that in many cases the background is itself non-stationary due to, for instance, fluctuations of instrumental noise or the presence of other transient sources in the field of view of the instrument (like pulsars). In these cases, to treat only the GRB contribution to the non-stationarity, we have to detrend the time series.
To do this, we fit the whole time series with a second order polynomial, taking into account only the portion of the data set far from the GRB prompt emission (i.e. the ending tail of the data series). We then remove the obtained polynomial background from the original data set.   

\subsection{DE statistics of the GRB catalog}

\begin{figure*}
\centering
\includegraphics[width=17cm]{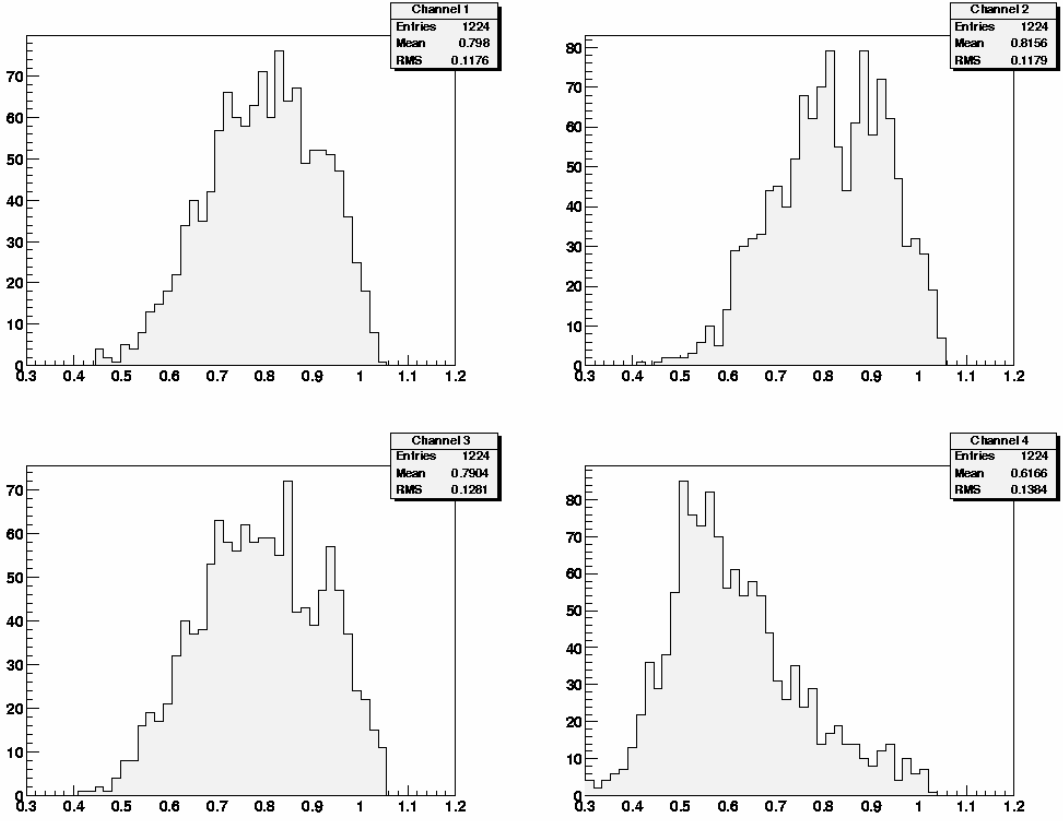}
\caption{Distribution of the Diffusion Entropy index over a sample of 1224 Long GRB calculated considering the first $2~T90$ portion of the light curves. Different panels correspond to different energy channels: 20-50 keV (Top-Left), 50-100 keV (Top-Right),100-300 keV (Bottom-Left), 300-1000 keV (Bottom-Right).
 DE indices mean and RMS values are shown in legends.}
\label{de_dist}
\end{figure*}

In the previous section we have shown how the complexity of a GRB can be characterized by the scaling index $\de$.
For GRBs we notice that the DE is correlated both with the intensity of the signal and with the duration of the prompt emission. 
The analysis of the entire catalog can give us an indication of how the two quantities (duration and intensity) affect the value of the $\de$ indices. 
The DE index scales  linearly with the $T90$ parameter. 
This trend is due to the balance between noise and GRB signal given by the ratio of the duration of the noise to the duration of the signal. 
More interesting is the relationship between the DE and the intensity of the signal. 
To remove the correlation with the duration of the signal, we select a portion of the light curve proportional to the the duration of the GRB. We cut the data series at twice the $T90$ value, to ensure that only the GRB signal is analyzed. Short bursts cannot be analyzed, since the portion of the light curve equal to 2 $T90$ is too short and the measurement of the scaling index $\de$ will be drastically affected by the lack of data. On the other hand, considering a longer portion of the light curve would exclude the longest bursts, for which the duration is similar to the duration of the data sets.
We then apply the DE analysis to the population of long GRB (1224 drawn from the BATSE catalog). The distributions of long bursts are shown in Fig.~\ref{de_dist} for the four energy channels.

The mean values of $\de$ for the whole catalog in the four energy bands are $0.80$, $0.82$, $0.79$, and $0.62$ respectively. The RMS are approximately the same in the four channels and are indicated in the upper right corner of each panel. The distributions are, in general, not symmetric, and the $\de$ indices are distributed in the range between 0.4 and 1. 
The mean scaling parameter $\bar\de$, and, more generally, the shape of the distributions are very similar for the first three channels, while $\bar\de$ is much smaller in the fourth channel. From this analysis, the first three channels seem to have the same degree of complexity, while the contribution of the non-stationarity/memory is, on average, weaker in the fourth channel. The number of counts in the highest energy range is small compared to the number of counts in channel 1,2, and 3. 
The case of completely correlated noise, also known as ballistic motion for which the walker always jumps in the same direction with a constant jump length, is the upper bound for the $\de$ value (=1). The distribution of the values in the first three channels indicates that the diffusive process is close to the ballistic case. This means that, in general, {\it GRB signals are highly correlated signals}.  

\subsection{The power-complexity relation}

\begin{figure}
\resizebox{\hsize}{!}{
\includegraphics{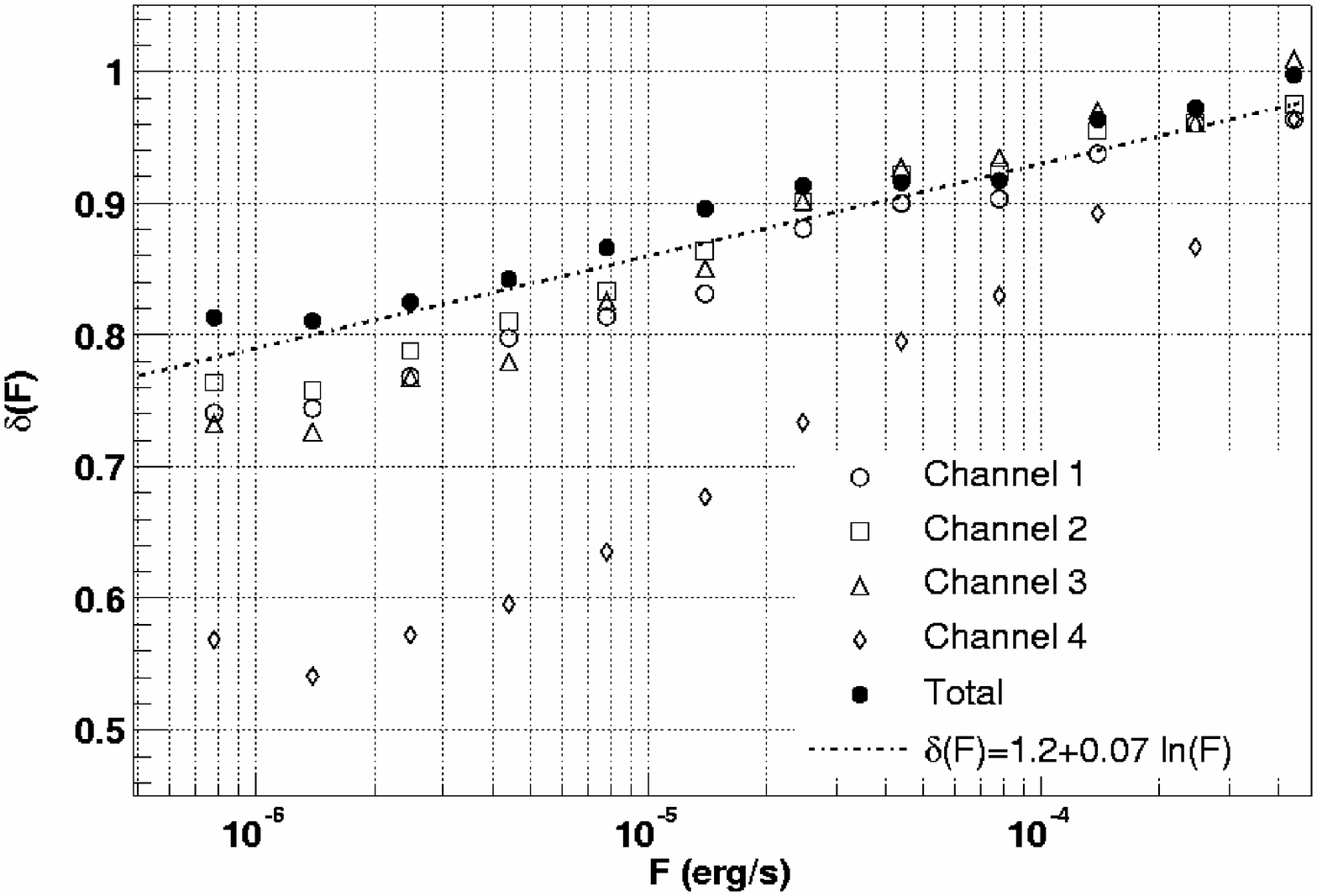}}
\caption{
{\it Power-Complexity} relation: the scaling index $\de(F)$, which is an estimator of the burst complexity as a function of the total fluence. Empty markers:  mean $\de(F)$  for each channel, as a function of the total fluence.
Fill dots: scaling index $\de(F)$ of the time series  resulting by summing the light curve in the four channels as a function of the total fluence.
The dash-dotted line is proportional to $A+B \log(F)$ with $A = 1.2$ and $B = 0.07$.
The scaling indices $\de(F)$ are averaged over fluence intervals to smooth statistical fluctuations. 
}
\label{de_f}
\end{figure}

In order to understand how the degree of complexity depends on the intensity, we perform another analysis. 
We consider the time series resulting from summing the time series in the four channels, using the sum of the fluences in the four channels as indicator of the intensity of the burst. We call $\de_{\mbox{tot}}$ the $\de$ index resulting by analyzing this time series.
In Fig.~\ref{de_f} it is shown the scaling parameter $\de_{\mbox{tot}}(F)$ of the light curve, as a function of the total fluence (filled dot) and the scaling parameter $\de_i(F)$ for the four different channels as a function of the total fluence.
Notice that the fluence is a measure of the energy emitted by the burst; thus, even if the count rate in the fourth channel is lower than in the first three channels, the fluences are similar. The plot shows that, at low fluences, the dominant contribution to the diffusion entropy is given by the first three channels where the non-stationarity of the light curve is already pronounced and the complexity is already developed. In the fourth channel the values of the $\de$ index are closer to $0.5$. As the fluence increases, the complexity in channel four increases and at $F\sim 10^{-4}\mbox{ erg/s}$ is comparable with the values of the first three channels.
The balance between the increment of the complexity with the fluence in the four channels results in the logarithmic relation found by fitting the $\de_{\mbox{tot}}$ as a function of the total fluence $F$.

We find a logarithmic dependence of the complexity of the data with respect to the fluence of the GRB: the higher the measured power, the greater the complexity of the GRB system.
We notice that this result qualitatively agrees with the variability- luminosity relationship reported by Fenimore \& Ramirez-Ruiz (\cite{feni00}), and Reichart et al. (\cite{reic01}). 
The fluence is a measure of the energy conversion efficiency in the GRB source, so the relationship between $\de$ and the fluence can be interpreted as the increase of the complexity of the energy conversion process.
This new relationship can be used as a testing ground for GRB scenarios or to constrain models, as, for instance, the internal shocks picture (Rees \& Meszaros \cite{rees94}, Kobayashi, Piran \& Sari \cite{kob97}, Daigne \& Mochkovitch \cite{daig98}, Piran \cite{piran99}), or the external shock scenario (Rees \& Meszaros \cite{rees92}, Chevalier \& Li \cite{chev99}, Dermer et al. \cite{derm95}).

\subsection{GRB simulations: a simple (Poisson) model}

\begin{figure}
\resizebox{\hsize}{!}{\includegraphics{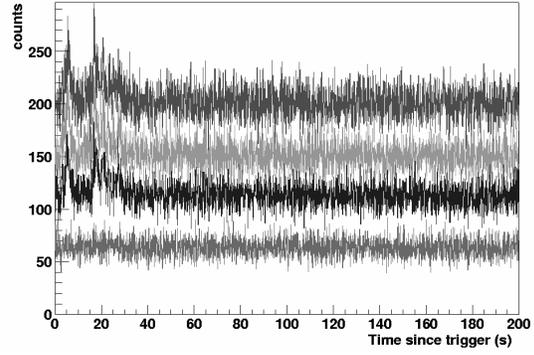}}
\resizebox{\hsize}{!}{\includegraphics{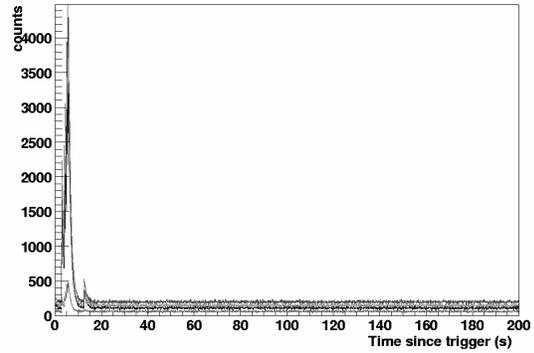}}
\caption{Light curves for two simulated burst. 
The light curves have been scaled in the four channels in order to take into account the different count rate in the different channels. Background noise has been added to the light curve.
Top panel:  The light curve is obtained using the peak shape of eq.~\ref{fred} with $A=30, N_{\mbox{p}}=30, t_{\mbox{d}}=1 s$. 
Bottom panel: $A=500, N_{\mbox{p}}=5, t_{\mbox{d}}=1 s$. The mean waiting time between pulses is $\tau=1$ for both the light curves.}
\label{synth1}
\end{figure}

We developed a schematic model by defining a pulse shape, and assuming that a GRB is made by the superposition of several pulses. 
We chose the simplest model by limiting the number of parameters, so we adopted a schematic fixed pulse shape: the simulated counts as a function of time are expressed by an exponential function:
\begin{equation}
F(t) =
\left\{
\begin{array}{ll}
 A \exp(-(t - t_{\mbox{p}})/t_{\mbox{d}}) & t \ge t_{\mbox{p}}\\
0 & \mbox{otherwise} \\
\end{array}
\right.
\label{fred}
\end{equation}
the pulse shape $F(t)$ depends on the value of $A$ which gives the amplitude of the spikes, on the peak time $t_{\mbox{p}}$ which gives the peak center, and on the time constant $t_{\mbox{d}}$. The duration of the spikes is given by the relation:
\begin{equation}
T90   = - t_{\mbox{d}} \ln 0.1, 
\label{fred duration}
\end{equation}
obtained by computing the temporal window which contains $90\%$ of the spike.
For a single spike, the value of the index $\de$ increases with $A$, and with the duration of the pulse. To reproduce the variability observed in GRB light curves we randomly extracted the amplitude of the spike (varied the value of $A$)  and we summed several peaks by extracting the waiting time between pulses from an exponential law with time constant $\tau$. 
To simulate analogous count rates we simulated the signal in the four channels, by taking into account the spectral shape of the typical GRB signal. 
In particular we used the mean values of the ratios of the fluences of the BATSE catalog in the different channels to normalize the simulated channels. 
We then added Poissonian noise to the synthetic light curves in order to simulate the background. 
What we obtained are plausible light curves. Each individual structure has a ``Fast Rise Exponential Decay'' (FRED) shape, in agreement with the observations. The count rates are comparable with those measured by BATSE, and the power density spectrum is a power law, in agreement with Beloborodov et al. (\cite{bel00}). In Fig.~\ref{synth1} are shown two typical light curves. 
In this model there are two time scales that determine the properties of the light curves. 
One is the duration of a single spike, which depends on the rise time and on the decay time. The second is the characteristic waiting time between pulses. 
These two characteristic times reflect the typical time scales at the sources for various physical models. 
For example, in the internal shocks scenario the rise and the decay times of the pulse are determined by the hydrodynamic time scale and by the angular spreading time scale (Sari \& Piran \cite{sari97}, Nakar \& Piran \cite{naka02}).
The separation between pulses reflects the activity of the ``central engine'' (Kobayashi et al. \cite{kob97}).  
In the external shock scenario, the variability of the light curve arises from the characteristic dimensions of the circumburst medium (Dermer \cite{derm99}). The duration of the peaks and the separation between consecutive pulses are related to the length scales of the irregularities of the surrounding medium. 
\begin{figure}
\resizebox{\hsize}{!}{
\includegraphics{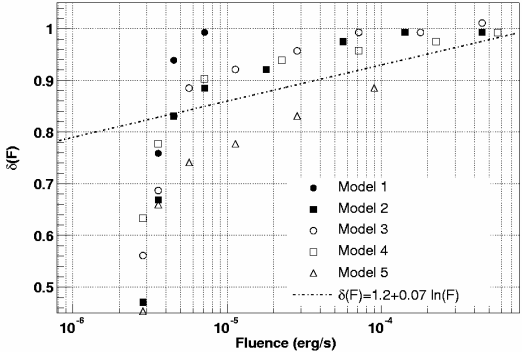}}
\caption{Simulated {\it power-complexity} relation. The synthetic light curves are obtained using a fred-like exponential peak shape. In the plot various relations have been obtained by varying the number of the peaks $N_{\mbox{p}}$, the waiting time between two consecutive peaks $\tau$, and the duration of the peaks $t_{\mbox{d}}$. 
Model 1: $N_{\mbox{p}}=50, \tau=10, t_{\mbox{d}}=1$. 
Model 2: $N_{\mbox{p}}=50, \tau=1, t_{\mbox{d}}=1$. 
Model 3: $N_{\mbox{p}}=50, \tau=0.1, t_{\mbox{d}}=1$. 
Model 4: $N_{\mbox{p}}=50, \tau=0.1, t_{\mbox{d}}=1$. 
Model 5: $N_{\mbox{p}}=100, \tau=0.1, t_{\mbox{d}}=0.1$. 
The dashed dotted line corresponds to the observed relations.
}
\label{models}
\end{figure}

In Fig.~\ref{models} we present the relations between the computed fluence and the scaling index $\de$ for synthetic light curves obtained from different combinations of the parameters.
Each curve is the result of 100 GRB, with fixed number of peaks $N_{\mbox{p}}$, fixed peak duration $t_{\mbox{d}}$ and fixed mean waiting time between peaks $\tau$, and varying the amplitude of the signal (by multiplying the light curve by a scale factor). The relation found analyzing the BATSE catalog, and plotted with the dashed dotted line, is the reference behavior for the ``intensity-complexity'' relation.

From this analysis we can see that if the characteristic time between pulses ($\tau$) is longer than the duration of the pulses ($t_{\mbox{d}}$), the growth of the scaling index $\de$ is too fast with respect to the observed logarithmic relation (model 1 in Fig.~\ref{models}). 
On the other hand, the solutions with an expectation time between the pulses shorter than the duration of the pulses follow better the observed relation.
This suggests that the preferred structure of the temporal series of GRB is composed by overlapped spikes while well-separated spikes make the $\de$ index growth too fast. Models 2, 3, and 4 follow, at high fluences, the observed relation. For these models the spikes are overlapping and the temporal duration of the structures is proportional to the number of peaks. Under the assumption that $t_{\mbox{d}}\ge \tau$ the different choices of the parameters give similar types of behavior. Model 5 shows that, if the number of the elementary spikes increases, the $\de$ index grows slowly, and reaches the upper limit at higher fluences.
At low fluences the growth of the index $\de$ for the simulated light curves is rapid compared to the observed relation. This is also because the DE method is very sensible to low signal-to-noise ratios, and, as soon as the fluence increases the light curve reveals its behavior in terms of complexity. 
This shows that a tuning between the number of the spikes, the duration of the spikes, and the time between spikes is necessary to obtain the observed dependence. 
The relation between the fluence and the scaling index $\de$ is a key study to understand the mechanism of production of gamma- ray and the environmental set up for GRB. 
 
\subsection{Diffusion Entropy as a function of time}

Our final application of the DE algorithm is the analysis of the time evolution of the $\de$ indices during a single GRB. This can be done by dividing the total light curve in overlapping intervals and applying the DE method to each of them.
We proceeded as follows: we divided the GRB light curve in $M$ overlapping windows; each of fixed size $W$. We applied the DE method to each new window separately. We chose overlapping windows to study the continuous time evolution of the $\de$ index. 
We refer to this procedure as the computation of the diffusion entropy index as a function of time (the reference time is the starting point of each series). 

\begin{figure}
\resizebox{\hsize}{!}{\includegraphics[width=10cm]{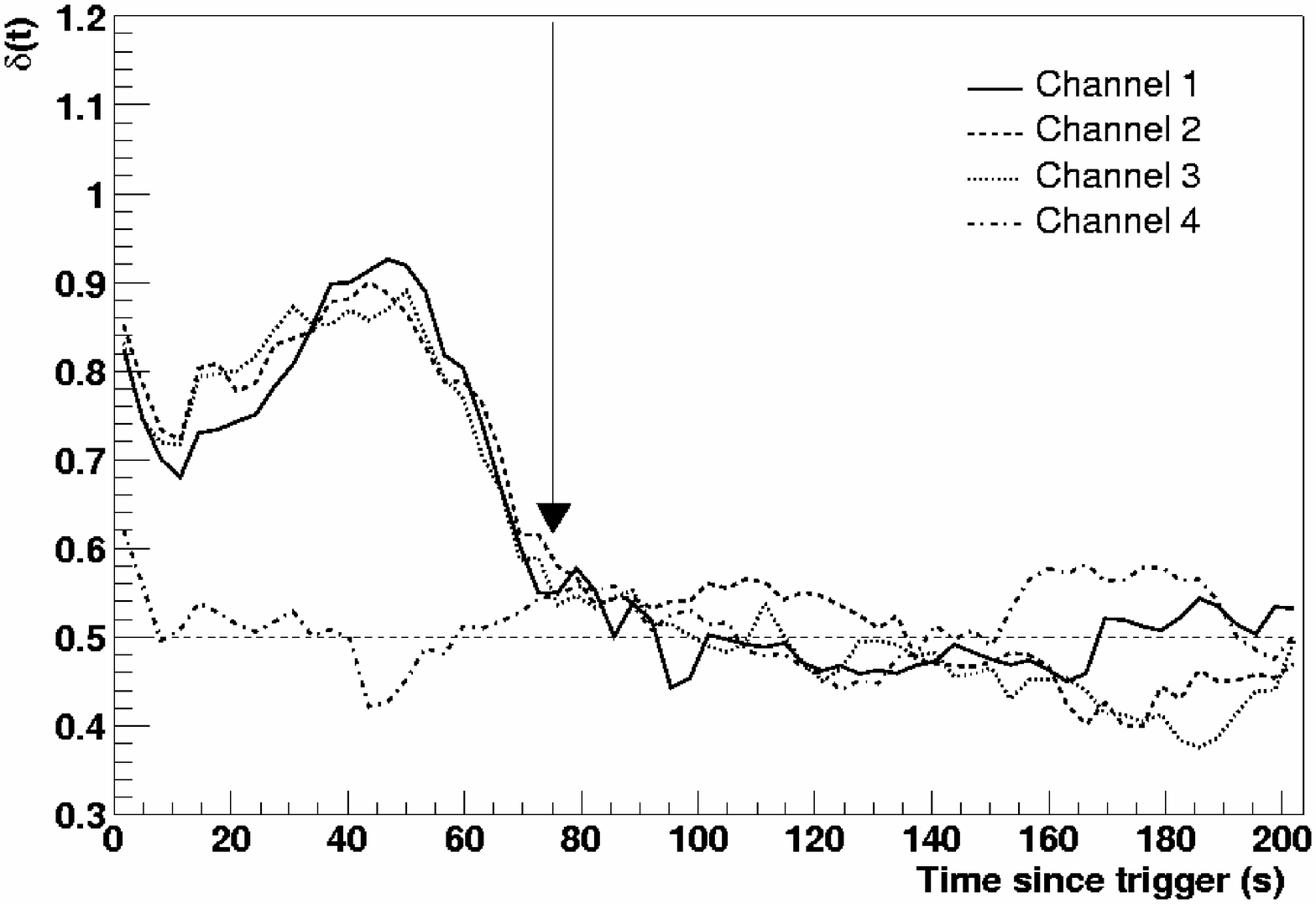}}
\resizebox{\hsize}{!}{\includegraphics[width=10cm]{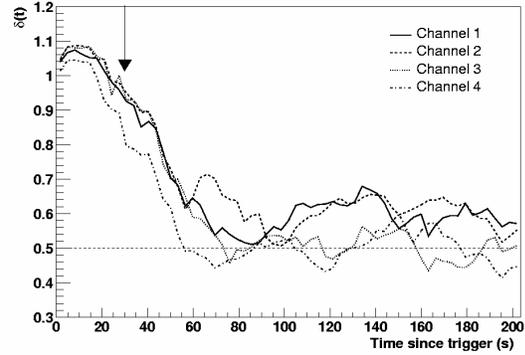}}
\caption{Diffusion Entropy as a function of time of GRB910429 (top), and GRB910601 (bottom). $W = 32$ seconds (500 points) and $M=0.64$ seconds (10 points).
Channel 1: 20-50 keV, Channel 2: 50-100 keV, Channel 3: 100-300 keV, and Channel 4: 300-1000 keV. Curves are averaged over $L=10$ realizations to smooth statistical fluctuations. The arrows indicate the end of the burst given by the $T90$ parameters.}
\label{grb_de_t}
\end{figure}

Figure~\ref{grb_de_t} shows the results of this analysis for the two selected bursts. The value of the scaling index $\de$  starts from one, indicating the presence of flaring activity, and then decreases, relaxing to the value 0.5 that indicates the end of the burst. 
The top panel of Fig.~\ref{grb_de_t} shows the behavior of the diffusion entropy index as a function of time for the GRB 910429. 
The activity of the burst is clearly visible in the first eighty seconds, in agreement with the recorded data of Fig.~\ref{grb_lc}. 
After 75 seconds the DE index reaches the dashed line at $\de_{\mbox{N}}$, representing the noise value. This estimate of the burst duration is in agreement with the $T90$ value for this burst.
The bottom panel of Fig.~\ref{grb_de_t} displays the DE index time evolution for GRB 910601.
The duration of the burst using the Diffusion entropy approach, is longer than the duration given by the $T90$ parameter. This is due to the fact that DE is sensitive to low signal-to-noise ratio, and the presence of a low signal after the main burst can keep the $\de$ index high. 
Furthermore, the scaling index $\de$  in the fourth energy channel approaches $\de_{\mbox {N}}$ more rapidly than in other channels. 
Using the simple model developed in the previous section we tested this behavior and we found that similar results can be achieved by  adding to the spiky simulated burst of Fig.~\ref{synth1} a synthetic peak with a long exponential decay time ($t_{\mbox{d}}\sim 50$) and small amplitude (comparable with the noise level). 
This artificial spike represents either a weak pulse, part of the burst main emission but longer, or a fluctuation of the instrumental background. For example, a saturation effect can cause a long-range memory effect.
 Since the fourth channel has a lower signal-to-noise ratio, the $\de$ index in this channel {\it systematically} reaches the noise value ($\de_{\mbox {N}}$) earlier than in the other channels, where  the signal-to-noise ratio is higher.
The DE method is sensitive to the balance between uncorrelated noise and non-stationary signal, and a shorter duration of the non-stationary signal in the fourth channel could be the cause of a lower $\bar\de$. This could explain why the mean $\de$ value for the GRB catalog in the fourth channel is slightly smaller than in the first three channels.

Moreover, the DE as a function of time shows that GRB 910601 seems to have a re-flaring from 90 to 140 seconds. 
In this time interval the $\de$ index grows again even if the signal intensity remains almost constant. 
This behavior is easy to see in Fig.~\ref{grb_de_t} (bottom) while it is almost invisible looking at the light curve (bottom panel of Fig.~\ref{grb_lc}). 
This shows that the DE could provide a new method to detect re-flaring activity and long time correlation in GRB time history. 
We have to stress that the presence of re-flaring activity has been observed in several cases in both short and long bursts.
This effect could, in principle, be considered when evaluating the duration of GRB, in particular in the case of short bursts when the prompt emission lasts less than two seconds but the anomalous activity seems to last longer. 
However, we notice that, in some cases, the increase of DE after the prompt phase is due to fluctuation of the background. 
Both the orbiting motion of the instrument and the presence of other flaring  sources in the field of view of the instrument can give an increment of the counts, and an increment of the $\de$ indices.
These cases have been confirmed by the BATSE catalog 
\footnote{http://gammaray.msfc.nasa.gov/batse/grb/catalog/4b/4br\_comments.html}. The DE approach detects these.

\section{Conclusions}

In this paper we performed a new analysis on the GRB phenomena applying a technique devoted to study the anomalous behavior in diffusive processes. The Diffusion Entropy method provides a measure of the non-stationarity/memory in GRB time series.
The scaling index $\de$ gives a well-defined value that represents the degree of complexity in a GRB light curve, with a precision of the order of few percent. In other words, the scaling index $\de$ estimates how far a GRB signal is from uncorrelated noise, for which $\de=\de_{\mbox {N}}=0.5$.
To remove the dependence of the scaling index $\de$ on the duration of the burst, and to obtain an estimation of the degree of complexity of the signal independently of its duration we selected a portion of the light curve proportional to the $T90$ value of the BATSE catalog. This also allows a general description of the GRB phenomena that does not depend on the particular experimental set up.  
The DE method can be successfully applied in the case of long bursts, while, in the case of short bursts, the analysis is more complicated due to the lack of statistics. 
 
For long bursts it has been possible to build up statistics applying the DE method on the BATSE catalog, obtaining a new description of the GRB phenomena in terms of their scaling index $\de$. 
The $\de$ indices are distributed between the value $\de_{\mbox {N}}$ and 1 with mean $\sim 0.8$. The mean values of the distributions of the $\de$ indices are almost the same for the first three energy channels, and smaller for the fourth channel.
Furthermore, using the time series obtained by summing the light curves in the four channels, we studied the dependence of the scaling index $\de$ on the measured power expressed by the total fluence. 
We found a logarithmic growth of the complexity with respect to the fluence. This feature connects a new variable (complexity) with an observable quantity (fluence), and the relationship between those quantities can be used to test GRB models and scenarios.

Using a simple statistical model where the parameters are the number and the duration of elementary spikes and the mean separation between them, we tried to reproduce the growth of the complexity with the fluence for synthetic bursts. 
We found that, in order to satisfy the observed relation between the $\de$ index and the fluence, the duration of the single spikes should not be shorter than the mean separation between spikes. 
The pulses of the synthetic light curves cannot be completely separated and the elementary structures have to be superimposed. 
The connection between the typical time scales of the signal and the typical dimensions at the source suggests a well defined scenario: the emitting regions are structured and the spikes of the light curve are the results of the superimposition of many elementary spikes.
 In the internal shock scenario, for example, the crossing time of the shells (duration of the spikes) has to be of the same order or even longer than the travel time between shocks (waiting time between pulses), which also reflects the activity of the central engine. 
In the external shock scenario, the variability is related to the irregularity of the surrounding environment and the sizes of the clouds have to be bigger than the separation between them.
 
To study the evolution of the complexity during a single Gamma-Ray Burst, we proposed a more detailed analysis by means of the application of the DE method to subset of data, obtained from the data by selecting overlapping windows.
In this case the DE method operates over smaller regions, compared to the global time history of the GRB: this implies that DE is more sensible to short time correlations. We found that the time that the $\de$ index takes to reach the $\de_{\mbox {N}}$ value (relaxation time) is shorter in the fourth energy channel compared with the relaxation time at lower energies. This behavior is due to the different signal-to-noise ratios of the different channels.
We showed that the DE index as a function of time permits the detection of weak flares (or re-flaring activities) after the prompt GRB emission. In several cases we found that the $\de$ index increased after the prompt phase. These spikes are weaker, compared to the prompt phase, and sometimes comparable with the background. They are indeed invisible looking at the light curve itself.  
The DE method is sensible to changes of statistics in the time series even if the intensity of the signal is comparable to the background level.

In further investigations we will use the available (first 1-2 seconds) high resolution (TTE) data for studying the short time correlation in GRB signal, especially in the case of short bursts, and to study the evolution of the scaling $\de$ index with higher temporal resolution. 
The Gamma-Ray Large Area Telescope (GLAST) will be able to collect more data and, thanks to its short dead time, it will record GRB signals from the deep universe with a temporal resolution never reached before. The light curves of the GRB will be resolved up to the millisecond time scale, and their structure will be better understood.

\begin{acknowledgements}
We are very grateful to Paolo Grigolini and Steve Shore for useful discussions during the preparation of this work.
\end{acknowledgements}

\end{document}